\documentclass[fleqn,twoside,twocolumn,nofootinbib,showkeys]{revtex4} % Specifies the document class %,unsortedaddress
\usepackage[sec,nocpr]{ujp} % \usepackage[cyr]{ujp} for cyrillic
\begin{document}
\title[Possible modification to standard cosmological model to resolve tension with Hubble constant values]%колонтитул
{POSSIBLE MODIFICATION TO STANDARD COSMOLOGICAL MODEL TO RESOLVE\\ TENSION WITH HUBBLE CONSTANT VALUES}%
\author{S.L.~Parnovsky}%1 автор
\affiliation{Astronomical Observatory of Taras Shevchenko Kyiv National University}%институт
\address{Observatorna str., 3, 04058, Kyiv, Ukraine}%адрес
\email{parnovsky@knu.ua}%e-mail

\udk{524.8} \pacs{95.30.Sf, 95.35.+d} \razd{\seci}

\autorcol{S.L.~Parnovsky}

\setcounter{page}{1}%

\begin{abstract}
The tensions between the values of Hubble constant obtained from the early and the late Universe data pose a significant challenge to 
modern cosmology. Possible modifications of the flat homogeneous isotropic cosmological $ \Lambda$CDM model are considered, in which 
the Universe contains dark energy, cold baryonic matter and dark matter. They are based on general relativity and satisfy two 
requirements:
(1) the value of the Hubble constant, calculated from the value of the Hubble parameter at the recombination by the formulas of flat $\Lambda$CDM model, should be equal to 92\% of the one based on low-reshift observations;
(2) deviations from the $\Lambda$CDM model should not lead to effects that contradict astronomical observations and estimations 
obtained thereof.
The analysis showed that there are few opportunities for choice. Either we should consider DM with negative pressure 
$-\rho_{dm}c^2\ll p_{dm}<0$, which weakly affects the evolution of the Universe and the observed manifestations of DM, or we should 
admit the mechanism of generation of new matter, for example, by a decay of DE.
\end{abstract}

\keywords{cosmology, Hubble constant tension,
dark matter, dark energy.}

\maketitle

\section{Introduction}\label{s1}

A great interest among cosmologists was caused by tensions between the values of the Hubble constant obtained from the observations 
of the early and the late Universe indicated recently in \cite{r}. 
I simply note that the estimation $H_0=67.4$ km s$^{-1}$ Mpc$^{-1}$ obtained from observations in the recombination era
account for about 92\% of the average of the estimations based on observations of not very distant objects $H_0=73.3$ km s$^{-1}$ Mpc$^{-1}$. 
The corresponding difference is at the level of $4\sigma-6\sigma$, which, according to \cite{r}, should be classified as something 
from a discrepancy or a problem to a crisis. They are robust to exclusion of any single method, team or source.

It is known that the expansion rate of the Universe is characterized by the time-dependant Hubble parameter $H$. Its current value 
is called the Hubble constant and is denoted by $H_0$. The value of the Hubble constant is calculated from the Hubble parameter 
measured in some era. This requires knowledge of both the characteristics of this era, usually its redshift $z$, and the cosmological 
model to relate these values. 

Estimations of the Hubble constant obtained by different methods are given in \cite{r}. Most measurements of the Hubble parameter 
occur at distances, which are small by cosmological standards. They have small redshifts and these measurements relate to the late 
Universe. However, a few measurements relate to the early Universe, more precisely to the recombination era (redshift 
$z=z_r\approx 1100$). 

First of all, these are CMB data from Planck satellite \cite{pl} and data from Dark Energy Survey Year 1 clustering combined with 
data on weak lensing, baryon acoustic oscillations and Big Bang nucleosynthesis \cite{abb}.

Naturally, the differences could be explained by measurement errors, including errors in observational data, their processing, or 
interpretation, or by an influence of some poorly accounted factors. That would 
mean cosmology is not yet an exact science as it pretends to be. For the purpose of this work I choose to ignore possible issues 
with observational data and consider them to be correct. This article demonstrates that the contradiction can be possibly eliminated 
by a modification of the standard $\Lambda$CDM cosmological model. It is clear that deviations from the $\Lambda$CDM model can 
not be large, because it is compatible with most observations.

I considered an isotropic homogeneous cosmological model in which the Universe is filled with dark energy (DE), dark matter (DM), 
and cold baryonic matter. The Universe has passed the era of recombination, so the influence of radiation and ultrarelativistic 
particles can be neglected. However, this model must be different from the $\Lambda$CDM model so that it can explain the 8\% 
difference in the Hubble constant at $z = z_r$ and $z = 0$.

I consider three possible modifications to it. In the first, 
the dark energy (DE) with an arbitrary equation of state is used instead of the cosmological constant $\Lambda$. In the second, 
the dark matter (DM) is not pressureless, although its pressure is small in comparison with the energy density. An example of such 
matter is the so-called warm dark matter (WDM).
A hypothetic WDM was introduced earlier in astrophysics to solve some problems associated with clustering on subgalactic scales 
and formation of halos \cite{bot}. WDM is often mentioned in explanations of a monochromatic signal 
around 3.5 keV in the spectrum of X-ray emissions from galaxy clusters like Perseus and Centaurus observed by XMM-Newton \cite{br,bm}.
The current most popular candidates for WDM particles are sterile neutrinos \cite{dh}, gravitinos, non-thermally 
produced WIMPs and other particles beyond the Standard Model.

The third modification is associated with the possibility of the transition or decay of DE into matter or vice versa.
The question is being investigated as to whether any of these modifications can explain the Hubble constant tension without 
contradicting other astronomical observations. Next step will be to consider in more detail options that meet this criterion.
 
\section{The choice of a cosmological model can affect the values of the Hubble constant}\label{s2}

I consider the flat relativistic 
isotropic homogeneous cosmological model with a scale factor $a(z)$ at the interval of redshift $z=a_0/a-1$ from the 
recombination era ($z=z_r$) to the present epoch ($z=0$). I assume that the Universe consists of dark energy (DE), 
cold baryonic matter, and dark matter (DM). An influence of radiation and ultrarelativistic 
particles can be neglected. The energy-momentum tensor of each component has a diagonal form $\mathbf{T}=\mathrm{diag}
(\rho c^2,p,p,p)$ in comoving frame. Note that I define the pressure of DM and DE just as a component of the 
corresponding energy-momentum tensor.

The Hubble parameter change law is described by the first Friedmann equation \cite{ll}
\begin{equation}\label{e1}
\begin{array}{l}
H^2=\frac{8\pi G}{3}\rho=\frac{8\pi G}{3}(\rho_m+\rho_{de})\\
=H_0^2\left(\frac{\rho_m}{\rho_{m0}}\Omega_{m0}+
\frac{\rho_{de}}{\rho_{de0}}\Omega_{de0}\right).
\end{array}
\end{equation}
Here $\rho$ and $\Omega$ are the density and the density parameter, the subscripts m and de denote matter and dark energy, the 
subscript 0 denotes the current value of the corresponding quantity, and $G$ is the gravitational constant. I assume that the 
Universe contains only matter, both dark and baryonic, and DE. This formula can also be obtained in the framework of non-relativistic 
cosmology \cite{pp}. 

The Hubble constant value is calculated from the Hubble parameter obtained by processing observational data 
by eq. (\ref{e1}) for the $\Lambda$CDM model. It assumes that the dark energy is the pure cosmological constant with constant density 
and both baryonic and cold dark matter are pressureless. For this model I use the standard subscript $\Lambda$ instead of de. 
Using the dependences of $\rho_{\Lambda}$=const and $\rho_m=\rho_{m0}(1+z)^3$ one get
\begin{equation}\label{e2}
H^2=H_0^2\left(\Omega_{\Lambda 0}+(1+z)^3\Omega_{m0}\right).
\end{equation}

According to the Planck satellite observations, the parameters of cosmological constant and matter density in the modern era are 
$\Omega_{\Lambda 0}=0.68 \pm 0.02$ and $\Omega_{m0}=0.32 \pm 0.02$ \cite{pl}, and their sum is fixed to 1 in the flat model. Note 
that these quantities are of the same order.

If the real cosmological model differed from the $\Lambda$CDM one, but we used (\ref{e2}) instead of (\ref{e1}), we would get not the 
value $H_0$, but rather the product $A(z)H_0$ 
\begin{equation}\label{e3}
\begin{array}{l}
H=H_0\sqrt{\frac{\rho_m}{\rho_{m0}}\Omega_{m0}+\frac{\rho_{de}}{\rho_{de0}}\Omega_{de0}}\\
= A(z)H_0\sqrt{\Omega_{\Lambda 0}+(1+z)^3\Omega_{m0}}.
\end{array}
\end{equation}   
with the factor 
\begin{equation}\label{e4}
A(z)=\sqrt{\frac{\frac{\rho_m}{\rho_{m0}}\Omega_{m0}+\frac{\rho_{de}}{\rho_{de0}}\Omega_{de0}}{\Omega_{\Lambda 0}+(1+z)^3\Omega_{m0}}}.
\end{equation}
   
If the $\Lambda$CDM model is correct we have $A(z)=1$. If it is not correct then $A(z)$ is almost equal to 1 for the late Universe but 
could differ from 1 for the early one. Let's try to explain the discrepancies in the values of the Hubble constant using Eq. (\ref{e4}). 
To explain the results of the article \cite{r} we need to provide $A(z_r)\approx 0.92$. It is clear that this means to go beyond the 
$\Lambda$CDM-model. We try to look for such modification which could provide the condition for $A(z_r)$ (the condition $A(0)=1$ is 
done automatically). An additional requirement is small deviations from the model, especially during the period of existence and 
evolution of galaxies and stars, i.e. from the reionization era or so-called ``cosmic dawn'', $z<z_g\approx 11$.

\section{DE cannot solve the problem}\label{s2de}

Let's start with the letter $\Lambda$ in the name of the model and consider dark energy with variable density instead of the 
cosmological constant. The CDM part remains the same. We get $A(z_r)<1$ if $\rho_{de}(z_r)<\rho_{de0}$, but we can not obtain 
$A(z_r)=0.92$. Indeed, 
\begin{equation}\label{e5}
A(z)>\sqrt{\frac{(1+z)^3\Omega_{m0}}{\Omega_{\Lambda 0}+(1+z)^3\Omega_{m0}}},
\end{equation}   
but this value is less than 1 by no more than $10^{-8}$ at $z\approx 1100$. This is caused by the value of the $(1+z)^3$ factor at large $z$. 
It's a dead end.

\section{DM with non-zero pressure}\label{s2a}

Now let's try to abandon the letters CDM in the model title and leave only the letters DM. This is not about the effects of 
electromagnetic radiation or 
ultrarelativistic particles and neutrinos. They exist, but their share is so small that even at $z\sim 1100$ they do not provide the 
desired value of $A(z)$. So temporally forget about them and suppose that matter is composed of ordinary matter and dark matter. 
We know enough about baryonic matter to be sure that it can be considered as pressureless one. Its density $\rho_b=\rho_{b0}(1+z)^3$. 

But we actually know just a bit about the DM. Suppose that it has some pressure $p_{dm}=w(z)\rho_{dm}c^2$ which can affect the evolution 
of its density $\rho=\rho_0F(z)(1+z)^3$. The first law of thermodynamics gives for
matter and DE (see \S~2.7.4 in \cite {pp})
\begin{equation}\label{e10}
\frac{d\rho}{\rho+p c^{-2}}=-\frac{dV}{V}=-3 \frac{da}{a}.
\end{equation}
Here $V\propto a^3$ is a volume of part of space expanding with the Universe, $a=a_0/(1+z)$ is the scale factor. 
This equation and corresponding EoS specifies the 
dependencies of densities of all components on $z$. I assume that each of the components, i.e. DE, baryonic and 
dark matter expands adiabatically. This means, in particular, that none of them could decay or transform into another. It is not 
difficult to find the relation 
\begin{equation}\label{e10a}
F(z)=\exp \left( 3 \int_{0}^{z}\frac{w(\xi)d\xi}{\xi+1} \right).
\end{equation}
If the density of matter $\rho_m$ is equal to the sum of the densities of baryonic $\rho_b$ and dark $\rho_{dm}$ matter, then from 
(\ref{e4}) we obtain the expression for $A(z)$ 
at large $z$ when it is possible to neglect the terms with subscripts de and $\Lambda$ 
\begin{equation}\label{e7}
A(z)=\sqrt{\frac{F(z)\Omega_{dm0}+\Omega_{b0}}{\Omega_{m0}}}.
\end{equation}
I used the value $\Omega_{b0}=0.16 \Omega_{m0}$ based on Planck data \cite{pl} to get the rough estimation $F(z_r)=0.82$ from the 
condition $A(z_r)=0.92$. From (\ref{e10a}) it is seen that this is impossible for $w(z)\geq 0$. That is, the DM must have a negative 
pressure on at least some time interval after recombination. So it cannot be called WDM.

\subsection{The case of the simplest EoS}\label{s2aa}

At first I consider the DM with the simplest and most popular equation of state (EoS) 
\begin{equation}\label{e6}
p_{dm}=w\rho_{dm}c^2,
\end{equation}   
where the subscript dm means dark matter, $\rho_{dm}$ and $p_{dm}$ are DM density and pressure, $c$ is the speed of light and $w=\mathrm{const}$. I 
calculate the parameter $w$ which could 
explain the difference in the estimated values of the Hubble constant obtained from high- and low-redshift data. From (\ref{e6}) 
and (\ref{e10}) we obtain the expression $F(z)=(1+z)^{3w}$. One get the rough estimation $w\approx -0.009$ from $F(z_r)=0.82$.  

It is clear that neither classical particles, nor bosons, nor fermions can have a negative pressure. So DM in this case should have a 
completely different nature than ordinary matter. On the other hand, the introduction of such a small negative pressure has little effect 
on the numerous astronomical manifestations of DM. I mean the rotation curves of galaxies, estimations of virial masses of galaxy 
clusters, gravitational lensing, galaxy cluster mergers like the Bullet Cluster (1E 0657-56) and so on (see Chap.~4 in \cite{pp}). 
It could affect the results of modelling of large-scale structure formation, however.

This pressure with $|w|\ll 1$ also weakly affects the evolution of the Universe. I will show that the changes are insignificant 
taking the age of the Universe as an example. It can be found from (\ref{e1}) for the flat model with cosmological constant, cold baryonic
matter, and DM with EoS (\ref{e6}). This age is equal to
\begin{equation}\label{t}
T=H_0^{-1}\int_0^1\frac{ u^{0.5+1.5w} du} {\sqrt{u^{3(1+w)}\Omega_{\Lambda 0}+u^{3w}\Omega_{b0}+\Omega_{dm0}}}.
\end{equation}
It is easy to calculate that at $w=-0.009$ it is 13.9 billion years or the 100.76\% of the age of the Universe for the $\Lambda$CDM model,
which is equal to 13.8 billion years. As one can see, the difference from the case of the $\Lambda$CDM model is negligible.

\subsection{More general EoS of DM} \label{s3}

Let us consider a more general EoS in the form
\begin{equation}\label{e8}
	w(z)= B(1+z)^\alpha,\; B<0,\alpha=const.
\end{equation}
Consider the $w$ change during evolution. It is easy to estimate the values of $w(z_r)$ for (\ref{e8}) with different $\alpha$.
At $\alpha\ge 0.1$ we get $w(z_r)\approx -0.066 \alpha$. At $\alpha<0$ we have $w(z_r)\approx 0.066 \alpha z_r^{\alpha}<0.066 \alpha $. 
Thus, for $\alpha <3$, the estimate gives $-0.2<w (z_r)<0$. This type of DM significantly distinguishes from DE because of $|w(z)|\ll 1$.

In the modern era, this parameter is equal to $w(0) = B$. For $\alpha> 0.2$ $B\approx -0.066 \alpha $, for $\alpha <-0.2$ 
$B\approx 0.066 \alpha$. Let`s estimate this parameter for ``cosmic dawn''. This corresponds to a redshift of $z_g \approx 11$.
At $\alpha \ge 0.2 $ we get $w(z_ g) \approx -0.066 \alpha \, 0.01^{\alpha}$, at $\alpha<0 $ we have 
$w(z_r)\approx 0.066 \alpha z_g^{\alpha}$. All these values are negative, but close to zero, so the pressure has little effect on 
the evolution of the universe and structures in it.

The exception is the very low values of $\alpha \ll -1$, which give a large value of $B$ in the modern era. At $\alpha \gg 1$ this 
problem arises in the recombination era. In the case $\alpha=1$ we have
\begin{equation}\label{e12}
\rho_{dm}=\rho_{dm0}(1+z)^{3}\exp(3Bz).
\end{equation}
Setting $F(z_r) \approx 0.82$ one can evaluate the combination
\begin{equation}\label{e14}
Bz_r\approx -0.066.
\end{equation}
This is the value of $w$ at the time of recombination. At present epoch we get $w = B \approx -0.066 / 1100 \approx -6\cdot 10 ^ {- 5}$ 
and the matter can be considered quite cold. The farthest known galaxy has $z\approx 11$. This corresponds to a value of 
$w \approx -0.0066$. Thus, galaxies appeared in the era when WDM was quite cold in the framework of the considered EoS. However, the 
structure began to form during the period of warmer dark matter.

Note that all obtained estimations for (\ref{e6}) and (\ref{e8}) lie in the parameter interval $-1/3<w<0$. Let me remind you that matter with $w>-1/3$ 
attracts surrounding bodies, but when $w<-1/3$, it repels them, demonstrating antigravity. This, in particular, is typical for the 
cosmological constant with $w = -1$ and DE with $w\approx -1$. However, astronomical observations show that DM attracts both ordinary 
matter and DM itself.

Naturally, we cannot use EoS (\ref{e8}) for $\alpha> 0$ and huge $z$ significantly exceeding $z_r$, since we come to the region with 
$w(z)<-1$. However, no one considers (\ref{e8}) as a real EoS, but only as a kind of approximation for the period after the recombination 
epoch. The evolution of the Universe before this epoch lies outside the framework of the considered model.
 
\section{Decay of DE into matter}\label{s3a}

Is it possible to provide the condition $F(z_r) = 0.82$ without invoking negative pressure? Suppose both baryonic and dark matter is presureless. 
Nevertheless, the density of matter is now $1:0.82 = 1.22$ times more than follows from the equation (\ref{e10}). In order to explain 
this, one could assume that its amount has increased from the recombination epoch. Since both types of matter have the same 
laws of density decreasing with time, both the decay of DM into ordinary matter and the transition of baryonic matter into dark matter 
cannot affect the total amount of matter. However, during the transition of DE into matter, an increase in its amount can be obtained 
without violating the laws of conservation of energy. 

So, one could suggest that the contents of the Universe is presureless, but its 
density has been decreasing since recombination more slowly than predicted by (\ref{e10}) due to the assumption that there is a source of 
influx of matter, dark or baryonic, through transition DE into matter. In addition, the process must be sufficiently intense so that from 
the moment of recombination the total amount of matter has increased by approximately 20\%.

This possibility is somewhat reminiscent of the now practically forgotten theory, proposed back in 1948 \cite{bg,h}. In it, matter was 
constantly born ``out of nothing'', more precisely from the mysterious C-field (C for creation), maintaining a constant density. Now we 
are sure that it is incorrect and the density of matter is constantly decreasing. However, it can be assumed that different components 
of the contents of the Universe can transform into each other, keeping its flatness. More specifically, DE can turn into matter.

\section{Conclusion} \label{s4}

Hubble constant tension can be explained in different ways, from errors in measurements, their processing and interpretation, to the 
manifestation of some unknown effects. But if we try to explain it, considering the $H_0$ values given in \cite{r} to be correct, 
and staying within the framework of cosmological models based on general relativity, in which the Universe contains DE, DM and 
cold baryonic matter, then we have few opportunities to choose from. Either we should consider DM with negative pressure 
$-\rho_{dm}c^2\ll p_{dm}<0$, which weakly affects the evolution of the Universe and the observed manifestations of DM, or we should 
admit the mechanism of generation of new matter, for example, by a decay of DE. None of the evolution of the DE density can explain 
the differences in the estimates given in \cite{r}.

Let me clarify that the mention of general relativity may suggest that the conclusions of the article are based on this theory. 
However, to obtain them, we needed two equations, namely (\ref{e1}) and (\ref{e10}). Each of them can be obtained within the framework 
of nonrelativistic cosmology (see \cite{pp}). Equation (\ref{e1}) is obtained from classical mechanics and Newtonian theory of universal 
gravitation (more precisely, the limit of general relativity for small curvature of space-time, generalizing this law), 
equation (\ref{e10}) from the first law of thermodynamics. Therefore, it is impossible to modify the homogeneous isotropic 
cosmological model by replacing general relativity with another theory, if this does not change the classical mechanics and thermodynamics.

It seems to me that the solution to the problem is more likely associated with the revision and refinement of the estimates of $H_0$ 
for the early and modern Universe, including taking into account the peculiarities of processing the initial data. A relevant analysis 
is presented in my later paper \cite{pufj}.
But in this article I focus on the deviations from the $\Lambda$CDM model and demonstrate that they theoretically could resolve the
differences in Hubble constant values obtained from high- and low-redshift observations with a negligible change to the 
age of the Universe and other parameters. However, they require the introduction of either a new effect, namely the possibility of DE 
transition into matter, or the assumption that DM has negative pressure, which excludes the possibility that it consists of currently 
known particles and even of practically all hypothetical particles considered as candidates for the role of DM.

\vskip3mm \textit{This work was supported by the National Research Foundation of Ukraine under Project No. 2020.02/0073. This paper was published 
as S.L. Parnovsky, Possible Modification of the Standard Cosmological Model to Resolve a Tension with Hubble Constant Values Ukr. J. Phys. 
Vol. 66 No. 9 P. 739-744 (2021). DOI: https://doi.org/10.15407/ujpe66.9.739}

\end{document}